\documentclass[a4paper,11pt]{article}
\pdfoutput=1
\usepackage[utf8x]{inputenc}

\usepackage{color,graphicx,dsfont,slashed,subfigure,multirow,enumitem,jheppub,amssymb}
\hypersetup{bookmarks=true,unicode=true,pdftoolbar=true,pdfmenubar=true,
	pdffitwindow=false,pdfstartview={FitH},pdftitle={Quantum-inspired event reconstruction with Tensor Networks: Matrix Product States},
	pdfauthor={J.~Araz,M.~Spannowsky}, pdfsubject={Tensor Networks},
	pdfcreator={J.~Araz,M.~Spannowsky},pdfproducer={texstudio}, pdfkeywords={Tensor Networks},
	pdfnewwindow=true,colorlinks=true,
	linkcolor=blue,citecolor=magenta,filecolor=magenta,urlcolor=cyan}
\definecolor{lightgrey}{gray}{0.9}

\def\btabu#1\etabu{\begin{tabular}{p{125mm}}#1\end{tabular}}
\def\btab#1\etab{\begin{tabular}{p{50mm}p{70mm}}#1\end{tabular}}
\def\btabnn#1\etabnn{\begin{tabular}{p{45mm}p{75mm}}#1\end{tabular}}
\def\btabx#1\etabx{\begin{tabular}{p{65mm}p{55mm}}#1\end{tabular}}
\def\btaby#1\etaby{\begin{tabular}{p{40mm}p{80mm}}#1\end{tabular}}
\def\btabyy#1\etabyy{\begin{tabular}{p{20mm}p{100mm}}#1\end{tabular}}
\def\btabzz#1\etabzz{\begin{tabular}{p{35mm}p{85mm}}#1\end{tabular}}
\def\btabyyy#1\etabyyy{\begin{tabular}{p{10mm}p{110mm}}#1\end{tabular}}
\def\btabyyyy#1\etabyyyy{\begin{tabular}{p{2mm}p{118mm}}#1\end{tabular}}
\def\btabwide#1\etabwide{\begin{tabular}{p{82mm}p{38mm}}#1\end{tabular}}
\def\bcen{\begin{center}}
	\def\ecen{\end{center}}
\def\bgfb#1\egfb{\bcen\fcolorbox{black}{lightgrey}{\parbox{130mm}{\btabu#1\etabu}}\ecen}
\def\bgfbn#1\egfbn{\bcen\fcolorbox{black}{lightgrey}{\parbox{130mm}{\btab#1\etab}}\ecen}
\def\bgfbnn#1\egfbnn{\bcen\fcolorbox{black}{lightgrey}{\parbox{130mm}{\btabnn#1\etabnn}}\ecen}

\def\bgfbx#1\egfbx{\bcen\fcolorbox{black}{lightgrey}{\parbox{130mm}{\btabx#1\etabx}}\ecen}
\def\bgfbyy#1\egfbyy{\bcen\fcolorbox{black}{lightgrey}{\parbox{130mm}{\btabyy#1\etabyy}}\ecen}
\def\bgfbyyyy#1\egfbyyyy{\bcen\fcolorbox{black}{lightgrey}{\parbox{130mm}{\btabyyyy#1\etabyyyy}}\ecen}
\def\bgfbzz#1\egfbzz{\bcen\fcolorbox{black}{lightgrey}{\parbox{130mm}{\btabzz#1\etabzz}}\ecen}
\def\bgfbyyy#1\egfbyyy{\bcen\fcolorbox{black}{lightgrey}{\parbox{130mm}{\btabyyy#1\etabyyy}}\ecen}

\def\bgfbalign#1\egfbalign{\bcen\fcolorbox{black}{lightgrey}{\parbox{130mm}{\btaby#1\etaby}}\ecen}
\def\bgfbwide#1\egfbwide{\bcen\fcolorbox{black}{lightgrey}{\parbox{130mm}{\btabwide#1\etabwide}}\ecen}

\newcommand{\be}{\begin{equation}}
\newcommand{\ee}{\end{equation}}
\def\bsp#1\esp{\begin{split}#1\end{split}}
\def\bpm{\begin{pmatrix}} 
	\def\epm{\end{pmatrix}} 
\renewcommand{\figureautorefname}{Fig.}

\def\sectionautorefname~#1\null{Sec.~#1\null}
\def\subsectionautorefname~#1\null{Sec.~#1\null}
\def\figureautorefname~#1\null{Fig.~#1\null}
\def\tableautorefname~#1\null{Table~#1\null}
\def\equationautorefname~#1\null{Eq.~(#1)\null}
\preprint{IPPP/20/114}

\begin{document}
	
	\date{\today}

	\title{Quantum--inspired event reconstruction with\\ Tensor Networks: Matrix Product States}
	\author[a]{Jack~Y.~Araz}
	\author[a]{and Michael~Spannowsky}
	\affiliation[a]{Institute for Particle Physics Phenomenology,\\Durham University, South Road, Durham, DH1 3LE,}

	\emailAdd{jack.araz@durham.ac.uk}
	\emailAdd{michael.spannowsky@durham.ac.uk}

	\vspace{10pt}
	\abstract{
	Tensor Networks are non-trivial representations of high-dimensional tensors, originally designed to describe quantum many-body systems. We show that Tensor Networks are ideal vehicles to connect quantum mechanical concepts to machine learning techniques, thereby facilitating an improved  interpretability of neural networks. This study presents the discrimination of top quark signal over QCD background processes using a Matrix Product State classifier. We show that entanglement entropy can be used to interpret what a network learns, which can be used to reduce the complexity of the network and feature space without loss of generality or performance. For the optimisation of the network, we compare the Density Matrix Renormalization Group (DMRG) algorithm to stochastic gradient descent (SGD) and propose a joined training algorithm to harness the explainability of DMRG with the efficiency of SGD.
	}

	%Use showkeys class option if keyword
	
	\maketitle

	%%%%%%%%%%%%%%%%%%%%%%%%%%%%%%%%%%%%%%%%%%%%%%%%%%%%%%%%%%%%%%%%
	\section{Introduction}\label{sec:intro}
	%%%%%%%%%%%%%%%%%%%%%%%%%%%%%%%%%%%%%%%%%%%%%%%%%%%%%%%%%%%%%%%%
	Over the last decades, Machine learning (ML) techniques have developed into standard tools for data analysis strategies in high-energy physics (HEP). Due to the requirement of analysing vast, highly correlated data in order to exploit the full physics potential of the LHC, it becomes more and more important to develop a fundamental understanding of the data analysis methods applied. Jet substructure analysis is a particularly popular research area where analytic reconstruction techniques~\cite{Kondo:1988yd,Soper:2012pb, Soper:2011cr, Soper:2014rya, Prestel:2019neg,Kasieczka:2020nyd} co-exist with numerical multivariate analyses methods~\cite{Brehmer:2018kdj,Brehmer:2019xox,Louppe:2016ylz, Kasieczka:2019dbj, Faucett:2020vbu}. The combination of a large amount of available data with an excellent theoretical understanding of the underlying physics in collider phenomenology provides the ideal environment to explore novel reconstruction techniques and to improve our understanding of existing approaches. 
	
	%	Whilst ML bestows fast and precise reconstruction; its application usually remains as a black box. Recently there has been a push towards understanding what ML techniques ``learns" from the data~\cite{Faucett:2020vbu}. \ms{mention explainable AI}
	
	A method of increasing popularity that is rooted in quantum mechanical concepts are tensor networks (TNs) or tensor network states~\cite{Bridgeman:2016dhh, Biamonte:2017dgr}. The amplitude of a wave function in quantum mechanics can be represented as a matrix for the superposition of multiple states where TNs comes into play to describe complex quantum many-body systems~\cite{Orus:2018dya, 10.5555/2011832.2011833}. TNs are \textsc{Lego}$^\circledR$-like constructions where the connection between each \textsc{Lego}$^\circledR$ piece represents the entanglement between two or more states.
	 %\ms{sentence here on quantum nature}\ja{already in the next sentence$ \to $}
	%TNs are \textsc{Lego}$^\circledR$-like construction to represent complex many-body quantum systems~\cite{Orus:2018dya}. Each connection between these ``\textsc{Lego}$^\circledR$" pieces represent entanglement between two or more local states. 
	A one-dimensional lattice, in this configuration, can be written as the Matrix Product States (MPS)~\cite{perez2006matrix, PhysRevLett.69.2863, PhysRevB.55.2164, PhysRevLett.93.040502,  PhysRevLett.91.147902} where more complex entanglements can be represented with tree tensor networks~\cite{Shi_2006, PhysRevB.82.205105} or multi-scale entanglement ansatz (MERA)~\cite{PhysRevB.79.144108, Vidal_2008}. It has been shown that tensor networks can be used to compress fully connected and convolutional networks to achieve more efficient results~\cite{garipov2016ultimate}. This study has been further expanded by using specialised MPS training techniques for image classification~\cite{stoudenmire2017supervised, novikov2017exponential, selvan2020tensor, efthymiou2019tensornetwork} and feature extraction~\cite{7207289}. MPS has also been used in unsupervised learning~\cite{Han_2018}, for anomaly detection~\cite{wang2020anomaly} and has been shown that it can produce comparable results to recurrent neural networks~\cite{xu2021tensortrain}\footnote{An extensive review on tensor networks in ML applications can be found in ref.~\cite{cichocki2014era}.}. Beyond MPS, there have been various ML applications with MERA~\cite{reyes2020multiscale, kong2021quantum} and in the 2D case with projected entangled pair states (PEPS)~\cite{Cheng_2021}. The transition to TNs also allows transferring the knowledge developed to understand and compute quantum many-body systems to ML. A particularly interesting feature of MPS is that the entanglement entropy of an MPS can shed light on the query of ``what the network is learning?"~\cite{martyn2020entanglement}.

	Despite the ever-growing interest in TNs, surprisingly little work has been dedicated to the applications of TNs in HEP, with the exception of ref.~\cite{trenti2020quantuminspired}. This study will show that MPS can be used to discriminate top jets over QCD jets with comparable precisions to state-of-the-art classifiers and that the tensor network learns the volume and correlations of the projected geometry of topological relations in the data, which is reflected by the entanglement entropy of the network. This observation can be exploited to reduce redundant information in the input data, thereby reducing the complexity of the network while maintaining a high classification performance. Thus, we propose different network pruning and optimisation techniques relying on the entanglement entropy of the system. 
	
	Traditionally the MPS has been optimized by dedicated algorithms such as Density Matrix Renormalization Group algorithm (DMRG)~\cite{Schollw_ck_2005, Schollw_ck_2011, Stoudenmire_2013, PhysRevLett.69.2863, McCulloch_2007, Bradley_2020} or Time Evolving Block Decimation (TEBD)~\cite{Vidal_2003,Vidal_2004}. These algorithms are designed to reduce the degrees of freedom in the given wave function to find the ground state energy of the acting Hamiltonian. Additionally, due to their construction, such algorithms allow the network to adapt to the complexity of the problem by expanding or reducing the connections between the local states. 
	Common neural networks, such as convolutional neural networks (CNN), recurrent neural networks (RNN) or deep neural networks (DNNs), are often optimised via stochastic gradient descent algorithms (SGD), which are known to be very effective methods to optimise a network. Such algorithms are designed to increase degrees of freedom in the network to map the given data on a higher-dimensional manifold. We will compare these two training methods and propose a novel way of combining the training algorithms to harness the adaptability and physical insight related to entropy entanglement of the MPS optimisation algorithm and the efficiency of SGD. We show that entanglement entropy calculated using DMRG can be used to identify redundant information in the data feature space and elucidate how the two training methods can be combined to one optimisation algorithm.
	
	This study has been organised as follows; in \autoref{sec:tns} we will give a detailed introduction into tensor networks  in mathematical form which will branch out into following subsections \ref{sec:mps} and \ref{sec:dmrg} where we will further discuss MPS and optimization algorithms. In \autoref{sec:toptagging} we will introduce our case study of top tagging and discuss our results. Finally we will conclude and summarize the study in \autoref{sec:conclusion}.
	
	%%%%%%%%%%%%%%%%%%%%%%%%%%%%%%%%%%%%%%%%%%%%%%%%%%%%%%%%%%%%%%%%
	\section{Tensor Networks}\label{sec:tns}
\begin{figure}[!h]
		\centering
		\includegraphics[scale=1.1]{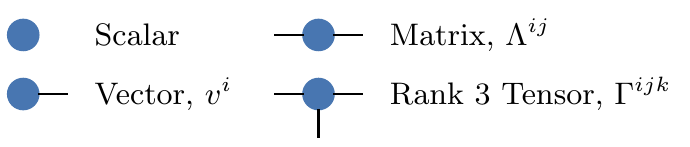}\\\vspace{5mm}
		\includegraphics[scale=1.1]{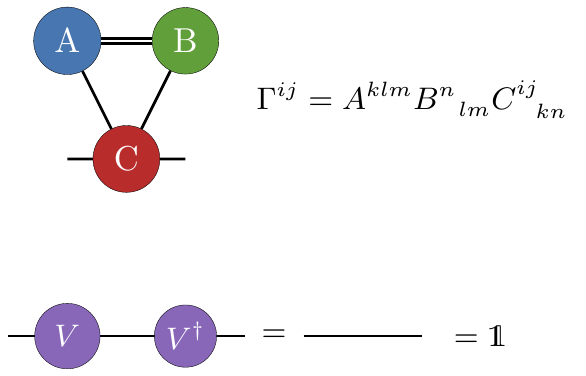}
		\caption{\it Top panel shows the Penrose representation of the basic tensor forms such as scalar, vector, matrix and a rank tree tensor. Bottom two  panels shows two examples of tensor operations where first notation shows tensor contraction between two rank tree ($ A^{klm} $ and $ B^n_{\ lm} $) and one rank four tensor ($ C^{ij}_{\ \ kn} $) resulting a rank two tensor ($\Gamma^{ij}$). The bottom equations shows a tensor contraction of a isometric tensor V which leads to identity.\label{fig:penrose}}
	\end{figure}

	%%%%%%%%%%%%%%%%%%%%%%%%%%%%%%%%%%%%%%%%%%%%%%%%%%%%%%%%%%%%%%%%
	Tensors are general multidimensional objects which can describe the multilinear relationship between algebraic objects within a vector space. For this study, we will use Penrose notation (or tensor diagram notation)~\cite{penrose1971applications} to describe tensors as briefly shown in \autoref{fig:penrose}\footnote{An extended review on tensor diagram notation and tensor networks can be found in refs.~\cite{Bridgeman:2016dhh,Orus:2013kga}.}. In this notation, a node without any edge describes a scalar, and each edge represents a higher rank object, such as one edge for vectors, two for matrices, where tensors can be rank N objects. The bottom two panels show two examples of algebraic operations with tensor diagram notation. Here, Einstein summation between two indices represented by connecting edges of nodes is assumed. Hence, the contraction between tensors $ A^{klm} $, $ B^n_{\ lm} $ and $ C^{ij}_{\ \ kn} $ has been shown by connecting edges with the same indices (for simplicity, indices are not shown in the figure). Similarly, an isometric matrix, $V $ is connected by its conjugate $V^\dagger$, which leads to Kronecker delta. Note that the Kronecker delta tensor is often shown as a line without any node. 

	Tensor networks are defined as a graph that describes the connection between several tensors into a composite tensor as shown in \autoref{fig:penrose} with $ \Gamma^{ij}$. The number of dangling edges describes the rank of this composite tensor. The Einstein summation has been applied to the connected edges where the leg between $A$ and $C$ indicates summation over index~$k$. These are called auxiliary (or bond) dimensions of the network. The size of these connections indicates each tensor's influence on each other, which will be further detailed in the following sections. Such objects have been widely used in the theoretical description of quantum many-body systems~\cite{Verstraete:2004cf}, and in the design of quantum computing algorithms.

	%%%%%%%%%%%%%%%%%%%%%%%%%%%%%%%%%%%%%%%%%%%%%%%%%%%%%%%%%%%%%%%%
	\subsection{Matrix Product States}\label{sec:mps}
	%%%%%%%%%%%%%%%%%%%%%%%%%%%%%%%%%%%%%%%%%%%%%%%%%%%%%%%%%%%%%%%%
	The Matrix Product States (MPS) or Tensor-Train (TT)~\cite{Fannes:1992vq, Klumper:1992vi, doi:10.1137/090752286, Bridgeman:2016dhh, 10.5555/2011832.2011833, Orus:2013kga} is one of the most studied tensor network topologies, widely used to describe 1D entangled quantum many-body systems~\cite{Orus:2013kga,10.5555/2011832.2011833, Verstraete_2006, Hastings:2007iok, Chen:2010gda, Schollw_ck_2011}. 

	\begin{figure}[!h]
	\centering
	\includegraphics[scale=1.1]{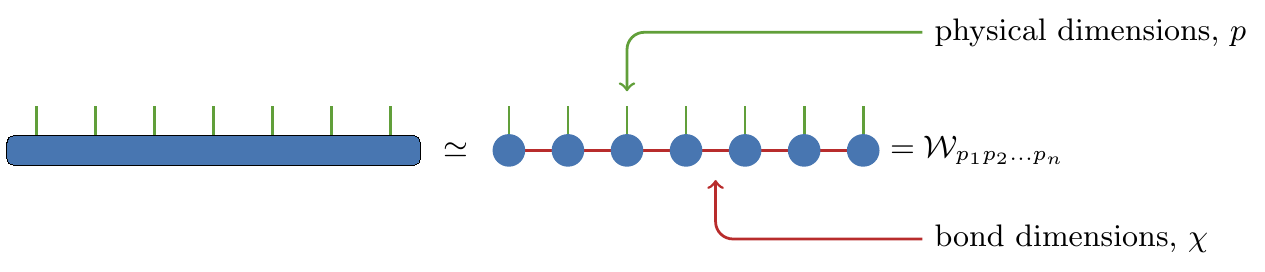}
	\caption{\it Rank seven tensor decomposed into Matrix Product States. Green lines represents the physical (Hilbert space) dimensions and red lines represents the auxiliary (bond) dimensions. \label{fig:full_mps}}
	\end{figure}
	
	The wave function for a general 1D lattice of N particles can be written as
	\begin{eqnarray}
		|\Psi\rangle = \sum_{p_1,\ldots, p_n=0} \mathcal{W}_{p_1\ldots p_n}\ | p_1 \rangle \otimes | p_2 \rangle \otimes \ldots \otimes | p_n \rangle \ ,  \label{eq:lattice}
	\end{eqnarray}
	where $|p_i\rangle$ are particle states spanning a local Hilbert space, $\mathcal{H}^{\otimes N}$, $ p_i $ denoting the position (or site) of the particles along the lattice and $\mathcal{W}_{p_1\ldots p_n}$ is rank-$ N $  amplitude tensor. The left-hand side of the equation shown in \autoref{fig:full_mps} represents a $\mathcal{W}_{p_1\ldots p_n}$ tensor in tensor-diagram notation. For simplicity, only seven out of $N$ edges are shown. Here the green edges represent the physical dimensions, $ p_i $. In this form, $\mathcal{W}_{p_1\ldots p_n}$ has $ \mathcal{O}(d^N) $ computational complexity where $ d $ denoting the number of indices that physical states can take, {\it e.g.} two for a spin state, $ |p_i\rangle \in \left\{ |\uparrow\rangle,|\downarrow\rangle \right\} $ or for particle state $ |p_i\rangle \in \left\{ |0 \rangle,|1\rangle \right\} $.
	
	For a classical, non-entangled system, $\mathcal{W}_{p_1\ldots p_n}$ can be factorized as $\mathcal{A}^{(1)}_{p_1}~\otimes~\mathcal{A}^{(2)}_{p_2}~\otimes~\ldots~\otimes~\mathcal{A}^{(n)}_{p_n}$ where local measurements on $\mathcal{W}_{p_1\ldots p_n}$ can be held independently. A locally entangled state, on the other hand, can be factorized as sum of products of amplitudes,
	\begin{eqnarray}
		\mathcal{W}_{p_1\ldots p_n} = \mathcal{A}^{\alpha_1}_{\quad p_1}\ \mathcal{A}^{\alpha_2}_{\quad\alpha_1 p_2}\ \mathcal{A}^{\alpha_3}_{\quad \alpha_2 p_3} \cdots \mathcal{A}^{\alpha_{n-1}}_{\quad \alpha_{n-2} p_{n-1}}\mathcal{A}_{\alpha_{n-1} p_n}\ .\label{eq:mps}
	\end{eqnarray}
	Here each tensor $ \mathcal{A} $ is connected to the one on the right and the left (except the first and last one). Although all the combinations are shown with $\mathcal{A}$, these tensors can be independent of each other. As before $p_i$ represents the physical states where $ \alpha_i $ shows the auxiliary indices between each tensor where the size of $ \alpha_i $ given as $ \chi $, as shown in the right-hand side of the equation in \autoref{fig:full_mps}. Such factorisation effectively compresses the computational complexity of the system to $\mathcal{O}(Nd\chi^2)$, assuming all the tensors, $\mathcal{A}$, have the same size of bond dimensions to the tensors on the left and the right. Note that this represents a state with open boundary conditions due to the fact that initial and final tensors are only connected to the tensor on the right and left, respectively. This structure ensures maximum entanglement between neighbouring sites, and entanglement decreases for the further sites depending on the size of the auxiliary dimensions. The entanglement between tensors is ensured via auxiliary dimensions where a larger bond dimension indicates a deeper influence of the node $\mathcal{A}_i$ in the network. However, the size of these ancilla dimensions also increases the computational cost; hence it needs to be adjusted with respect to the necessary accuracy required from the system. Note that in MPS, the network's topology is restricted at one dimension; hence the influence of a feature can only be observed on the features on their right and left.
	
	The amplitude tensor can be unfolded to the form in \autoref{eq:mps} via Singular Value Decomposition (SVD)~\cite{10.1093/qmath/11.1.50, Eckart:1936va}. Each physical dimension, $p_i$ can be splitted as
	\begin{eqnarray}
		\mathcal{W}_{p_1\ldots p_n} & =& U^{\alpha_{n-1}}_{\quad p_1\ldots p_{n-1}}\ \underbrace{S_{\alpha_{n-1}\beta}\ V^\beta_{\ \ p_n}}_{\mathcal{A}_{\alpha_{n-1} p_n}} \nonumber \\
		& =& U^{\alpha_{n-2}}_{\quad p_1\ldots p_{n-2}}\ \underbrace{S_{\alpha_{n-2}\beta}\ V^{\beta\alpha_{n-1}}_{\quad\ p_{n-1}}}_{\mathcal{A}^{\alpha_{n-1}}_{\quad \alpha_{n-2} p_{n-1}}} \ \mathcal{A}_{\alpha_{n-1} p_n} \nonumber \\ 
		& \vdots &  \label{eq:svd}\\
		& = & \underbrace{U^{\alpha_1}_{\quad p_1}}_{\mathcal{A}^{\alpha_1}_{\quad p_1}}\ \underbrace{S_{\alpha_1\beta} V^{\beta\alpha_2}_{\quad\ p_2}}_{\mathcal{A}^{\alpha_2}_{\quad\alpha_1 p_2}} \ \cdots\ \mathcal{A}^{\alpha_{n-1}}_{\quad \alpha_{n-2} p_{n-1}}\mathcal{A}_{\alpha_{n-1} p_n}\ , \nonumber 
	\end{eqnarray}
	where $U$ and $V$ are isometric tensors ($UU^\dagger=VV^\dagger = \mathds{1}$) with respect to the decomposition index and $S$ are diagonal tensors of singular values per for factorization of Hilbert space dimensions. It is important to note that this decomposition is not unique. One can immediately observe that we could start the decomposition from the right-most leg of the $\mathcal{W}$ tensor and combine $S$ tensors with $V$ to form $\mathcal{A}$. Equally, one could have started from the left-most tensor and absorb $S$ with $U$ to form $\mathcal{A}$, which would give the same result after contraction as \autoref{eq:svd}. Similarly, leaving $S$ unabsorbed would not change the outcome either. Hence each $S$-tensor corresponds to gauge choices~\cite{10.5555/2011832.2011833}.
	Although SVD splits amplitude into bitesize tensors, it is still quite expensive to do algebra with such an object for a large bond dimension. Trimming the singular values below a certain threshold, $\epsilon$, allows using significantly smaller tensors. However, one pays the price of a reduced precision in the approximation of the full tensor.	
	Recently, it has been shown that a fully connected neural network can be expressed via a tensor network~\cite{novikov2015tensorizing} and even a convolutional network can be compressed into a tensor network structure~\cite{garipov2016ultimate}. A neural network defines an affine transformation between the feature space and the output of the layer, 
	\begin{eqnarray}
		f^l_{FC}(\mathbf{x}) =  \sigma_2\left( \sigma_1\left(\mathcal{W}^{(1),m}_{\ i}\ \Phi^i(\mathbf{x^{(n)}}) + \mathcal{B}^{(1),m}\right)^m \mathcal{W}^{(2),l}_{\ m} + \mathcal{B}^{(2),l}\right)^l\ , \label{eq:dnn}
	\end{eqnarray}
	where $\mathcal{W}$ and $\mathcal{B}$ are weight and bias tensors, respectively and $\Phi(\mathbf{x})$ is the feature tensor.  $ f^l_{FC}(\mathbf{x}) $ denotes outputs of the network\footnote{FC stands for fully connected network.} where it can be further transformed and $ \sigma_i $ are the activation functions. Each transformation refers to a layer which aims to find a different vector-space to represent the previous layer or input. Ref.~\cite{novikov2015tensorizing} shows that compressing weight tensor via an MPS can drastically improve the performance by increasing the ability to represent the feature space with a single linear transformation instead of interconnected hidden transformations. Since the MPS asserts a linear relation between the features, it also increases the interpretability of the network.
	
	In order to benefit from an MPS layer, one first needs to map the feature tensor in a tensor product form~\cite{stoudenmire2017supervised, novikov2017exponential},
	\begin{eqnarray}
		\Phi^{p_1\cdots p_n}(\mathbf{x}) = \phi^{p_1}(x_1)\otimes \phi^{p_2}(x_2)\otimes \cdots \otimes\phi^{p_n}(x_n)\quad ; \quad \phi^{p_i}(x_i) = \alpha|0\rangle \oplus  \beta|1\rangle \ , \label{eq:feature_map}
	\end{eqnarray}
	where $\phi^{p_i}(x_i)$ shows the mapping of each feature $x_i$ up to coefficients $ \alpha $ and $ \beta $. Note that \autoref{eq:dnn} requires a feature vector but in \autoref{eq:feature_map}, we have a rank-N feature tensor where N is the number of features. There have been several proposals for feature mapping where ref.~\cite{stoudenmire2017supervised} discusses a trigonometric mapping, $\phi^{p_i}(x_i) = [\cos(x_i \pi k),\ \sin(x_i \pi k)]$, and ref.~\cite{novikov2017exponential} discusses a polynomial mapping, $\phi^{p_i}(x_i) = [1,\ x_ik]$, of the feature-vector\footnote{$ k $ stands for an arbitrary scaling constant.}.
	
	\begin{figure}[!h]
		\centering
		\includegraphics[scale=1.2]{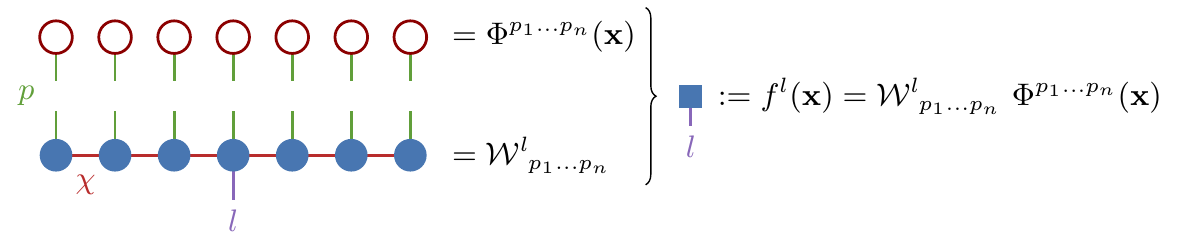}
		\caption{\it Feature tensor contraction with an MPS. Red nodes represents each mapped feature, blue nodes represent individual tensors of an MPS, the green, red and purple line represents physical dimensions, $ p_i $, auxiliary dimensions, $ \chi $ and output label, $ l $.  \label{fig:mps}}
	\end{figure}
	After the mapping, an MPS can be connected to each of the physical dimensions denoted by $p_i$. \autoref{fig:mps} shows the contraction between $\Phi^{p_1\cdots p_n}(\mathbf{x}) $ and the MPS where red circles represent the outer product of the feature-tensor, and blue circles represent the MPS. The contracted tensor results in a rank-1 tensor (vector) $f^{l}(\mathbf{x}^{i})$ for a single example. The Born rule dictates that the square of a wave function is the probability of the measurement; hence for this study, we will use $|f^{l}(\mathbf{x}^i)|^2$ as the prediction probability of the network.
	%\ms{dont understand this sentence: Contraction results with the decision made by the network where the dimensionality of the output represented by $l$, which depends on the classes required from the network.} 
	Note that the auxiliary dimensions, $\chi$, holds the information regarding the entanglement between each site. For MPS, the entanglement defines the correlation of a site to the block on the left or right. As discussed before, although the size of the bond increases the computational complexity, it also enables each site to be entangled with further away sites rather than only the neighbouring sites. The influence of each bond can be measured by its entanglement entropy~\cite{martyn2020entanglement}.

	Using an MPS for classification also allows us to use the quantum theory built around MPS. Measuring the entanglement entropy on the MPS nodes allows us to interpret the correlation between two neighbouring features, and using this information, one can further adjust the network size or feature space mapping to achieve more efficient classification~\cite{Orus:2018dya}. Using the Schmidt decomposition, one can write a bipartite quantum state via its orthonormal basis. The Schmidt decomposition can be achieved via SVD for the centre-of-orthogonality tensor by separating the eigenvalues from the node,
	\begin{eqnarray}
		| \Psi\rangle_{XY} = \sum_{i}^{\chi} \lambda_i |U\rangle_X |V\rangle_Y\ . \label{eq:schmidt_decomp}
	\end{eqnarray}
	Here $\lambda_{i}$ are positive definite singular values (Schmidt coefficients) in the $S$ tensor, and as before $|U\rangle_X, |V\rangle_Y$ is the orthonormal basis (Schmidt basis). The entanglement entropy then can be calculated via the von Neumann entropy~\cite{vonneumann},
	\begin{eqnarray}
		\mathcal{S} = - \sum_{i}^{\chi} \lambda_i^2 \log \lambda_i^2\ . \label{eq:entanglement_entropy}
	\end{eqnarray}
	The entanglement entropy indicates how strongly two sites are connected, where zero means the two sites decouple. The entropy is bounded from above by the area law where for each site $\mathcal{S} \leq \log\chi$~\cite{Eisert:2008ur}. The normalized singular value tensor, $S$, has all the singular values in descending order where $\lambda_1 = 1$. If there is no other singular value but $\lambda_1$, the entropy of the node will be zero, which means that it does not hold valuable information for classification~\cite{Wang:2012vn}. As will be demonstrated in \autoref{sec:training}, if neighbouring sites have similar entropy, only the one with largest entropy can be kept to shrink the feature-space without loss of generality. Whilst this gives extensive interpretability of the network, it also allows us to optimize the network by eliminating redundant features. 
	
	\begin{figure}[!h]
		\centering
		\includegraphics[scale=1]{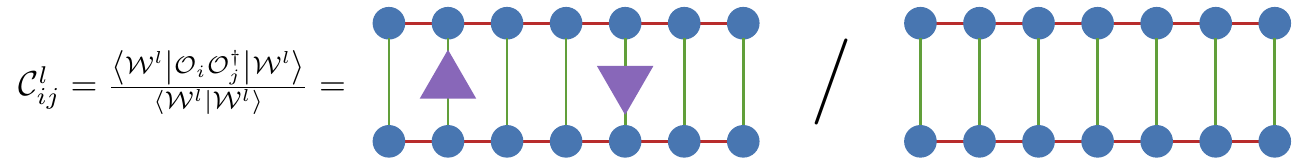}
		\caption{\it Two site correlation of the MPS, shown with blue nodes and $|\mathcal{W}\rangle$, with respect to an operator, $\mathcal{O}$ also represented via purple triangles. $i(j)$ represents the location of the sites to which the operator has been applied, and $l$ is the label index.} \label{fig:mps_corr}
	\end{figure}
	Beyond entanglement entropy, two site correlations of the MPS can be calculated depending on an operator. This operator can be chosen from the particular group that the data embedding is based on; for instance, the spin states are based on the $SU(2)$ group, and the generators of this group are Pauli matrices. \autoref{fig:mps_corr} shows the two-site normalised correlator where the operator has been shown by $\mathcal{O}$ and $i(j)$ represents the sites that the operator has attached. The right-hand side of the equation shows the same in Tensor Diagram notation, where purple triangles represent the operator and its Hermitian conjugate. $C^{l}_{ij} = 1\ (-1)$ denotes fully (anti-)correlated sites on a given basis, where one can discard one or the other site without loss of generality. $C^{l}_{ij} =0$, on the other hand, indicates no correlation, which means that both sites bring valuable information to the process at hand. As before, $l$ indicates the prediction label; hence the correlation can be calculated with respect to signal or background. This property of TNs has also been exploited in ref.~\cite{trenti2020quantuminspired}.
	
	MPS have been optimized using dedicated algorithms like the DMRG~\cite{Schollw_ck_2005, Schollw_ck_2011, Stoudenmire_2013, PhysRevLett.69.2863, McCulloch_2007, Bradley_2020}, TEBD~\cite{Vidal_2003,Vidal_2004} and recently, it has been shown that usage of alternating least squares methods in MPS minimization can be quite efficient~\cite{doi:10.1137/100818893}. All these methods are designed to contract a given Hamiltonian with a 1D lattice wave-function, shown in \autoref{eq:lattice}, and find the ground state energy efficiently. Ref.~\cite{stoudenmire2017supervised} shows that it is possible to use DMRG-like updating algorithms for classification tasks as well, which will be discussed in \autoref{sec:dmrg}. However, neural networks traditionally updated using SGD-based backpropagation algorithm, which can also be used for MPS type of networks~\cite{efthymiou2019tensornetwork}. 
	
	\begin{figure}[!h]
	\centering
	\includegraphics[scale=1]{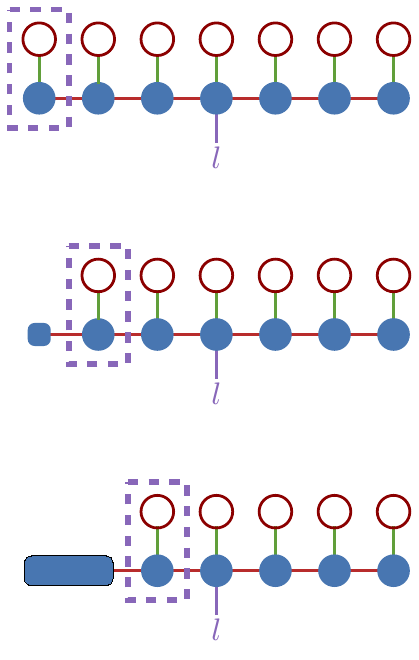}
	\caption{\it Schematic representation of the contraction sequence of MPS, dashed line shows the contraction with the feature space at each step and rectangular node represents the vertical contractions. The figure evolves from top to bottom.} \label{fig:keras-like}
	\end{figure}
	
	For the SGD-based backpropagation, an efficient contraction method is needed. Since all the components of the network cannot be contracted at once, one needs to use the so-called bubbling method~\cite{Bridgeman:2016dhh} to contract each site. \autoref{fig:keras-like} shows the bubbling sequence for such contraction.  As before, red circles represent the training sample preprocessed in the form of \autoref{eq:feature_map}, and the blue circles represent the MPS. The sequence evolves from the top frame to the bottom, as shown with the dashed enclosing line, where the first site on the left contracted with the first MPS node, then in the second step the second MPS node is contracted with the second site. In the third step, while the third site has been contracted with the third node, two previous contractions have been merged. This sequence continues until all the tensors are contracted, and the final $f^l(\mathbf{x})$ have been reached\footnote{Similar contraction methods have been adapted in refs.~\cite{efthymiou2019tensornetwork, torchmps}.}. Note that, although the label edge is placed at the centre of the MPS in \autoref{fig:keras-like}, the replacement of it will not affect the result; as a matter of fact, such bubbling algorithm will be most efficient if the label edge is placed at the rightmost tensor. This will allow the contraction order to be constant until the last node. Once the full contraction is complete one can calculate the loss function and each tensor can be updated via a specific SGD algorithm such as \texttt{Adam}~\cite{Kingma2014AdamAM}.

	Whilst such an approach leads to a very efficient algorithm, it lacks the adaptation features of the DMRG algorithm. It has been shown that a DMRG-like updating algorithm allows a network to automatically adjust its bond dimension for the complexity of the problem~\cite{stoudenmire2017supervised}. In \autoref{sec:dmrg} we will discuss how such updating scheme is possible.
	
	%%%%%%%%%%%%%%%%%%%%%%%%%%%%%%%%%%%%%%%%%%%%%%%%%%%%%%%%%%%%%%%%
	\subsection{Learning through Density Matrix Renormalization Group Algorithm}\label{sec:dmrg}
	%%%%%%%%%%%%%%%%%%%%%%%%%%%%%%%%%%%%%%%%%%%%%%%%%%%%%%%%%%%%%%%%
	
	A DMRG-like MPS updating algorithm for classification purposes was first proposed in ref.~\cite{stoudenmire2017supervised}. In this algorithm, tensors are updated two-by-two for the reasons that will be shown below. Originally, the DMRG algorithm is used to find the ground state energy of a  Hamiltonian acting on a 1D lattice where the algorithm sweeps the lattice from one side to the other and optimizes the energy by updating the site its currently on. It iteratively reduces effective degrees of freedom to emphasize the most prominent features. The same principle can be applied to classification tasks. 
	
	\begin{figure}[!h]
	%\hspace{-7mm}
	\centering
	\includegraphics[scale=.95]{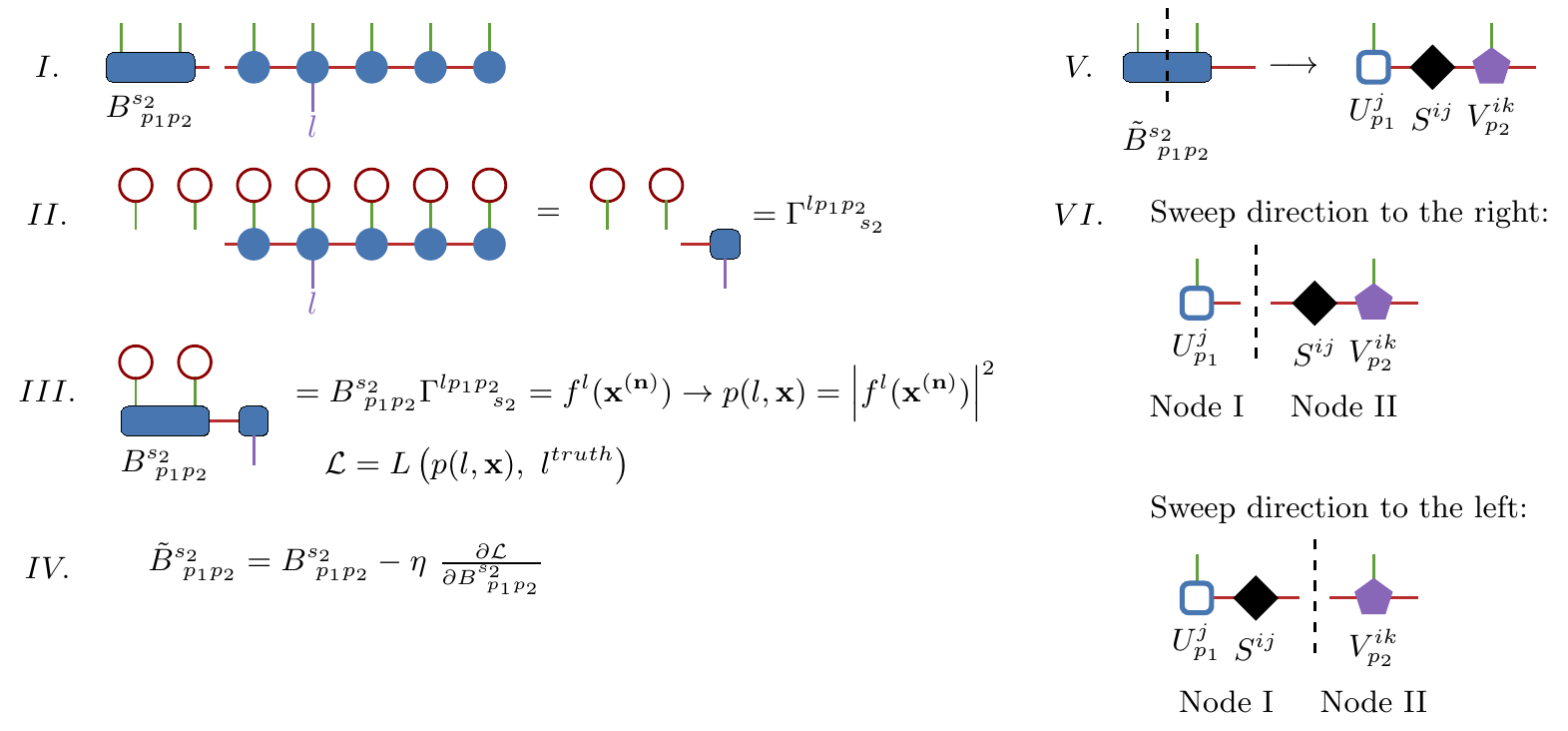}
	\caption{\it Schematic representation of two-site DMRG-like algorithm used for updating tensor network. I - III shows the forward-pass where the loss function is calculated and IV - VI shows the backpropagation where two nodes of the MPS has been updated using gradient decent and splitted back into its original two node structure via SVD. \label{fig:dmrg}}
	\end{figure}

	\autoref{fig:dmrg} shows the step-by-step evolution of a DMRG-like updating algorithm where I-III represents the forward pass, and IV-VI represents the backpropagation portion of the algorithm. First, let us assume that we have a network in the form of \autoref{fig:mps}. In order to update two nodes, one needs to calculate the gradient of the loss function with respect to the contraction of these two nodes. In step I, we calculate the contraction of these two tensors resulting in a rank-3 tensor, $B^{s_2}_{\ p_1p_2}$. For the sake of simplicity, this procedure has only been shown for first two tensors. Here, $s_2$ represents the auxiliary leg between the second node and the rest of the MPS. Then, in step II, we contract the rest of the MPS with $\Phi^{p_3\cdots p_n}(\mathbf{x}^{(i)}) $ via the bubbling algorithm shown in \autoref{fig:keras-like} and take the outer product with the two sites that are not connected to the MPS, resulting in a rank-4 tensor, $\Gamma^{ls_2}_{\ p_1p_2}$. Here, $i$ represents a particular training example, where each training example goes through the same process. Finally, on step III, the prediction, $p(l,\mathbf{x})$, is calculated by contracting $\Gamma^{ls_2}_{\ p_1p_2}$ with $B^{s_2}_{\ p_1p_2}$ and the objective function, $\mathcal{L}$, can be calculated using the entire training batch, which completes the forward pass of the algorithm. It is important to note that, in this study, we will use the tensor network as a Born Machine, where the square of the state, {\it i.e.} the wave-function, corresponds to the probability of the prediction.
	
	For the backpropagation portion, one first needs to calculate the gradient of the loss function with respect to $B^{s_2}_{\ p_1p_2}$ and the contracted tensor can be updated via gradient descend, shown in \autoref{fig:dmrg} panel IV. Here $\eta$ stands for the learning rate, which controls the step size of the gradient descent. Choosing a small learning rate may cause a prolonged convergence to the desired outcome; a large learning rate, on the other hand, may lead to fast convergence but can skip over the global minima. Hence, its value needs to be adjusted throughout the training. 
	
	In the following, the updated tensor, $\tilde{B}^{s_2}_{\ p_1p_2}$, is decomposed using SVD  to replace the initial two nodes. This procedure continues step by step until the last tensor on the right is updated. Then, the algorithm starts this time from the right side and applies the same algorithm until it reaches the starting position. As discussed earlier for \autoref{eq:svd}, there is not a unique way of contracting the singular values, $S$, with the eigenvector tensors. As shown in \autoref{fig:dmrg} panel VI, we choose to contract $S$ to the tensor on the right while sweeping is done towards the right direction, and $S$ are contracted with the tensor on the left for the left sweeping direction.
	
	As a virtue of SVD, the size of the auxiliary dimensions is modified during each update. This allows the network to increase the reach of the entanglement of each site, effectively increasing the entanglement entropy. Hence, the network grows automatically without the need to explicitly determining the size of each bond. However, an upper limit to the number of allowed singular values is necessary; otherwise, the network will grow to an unmanageable scope for the computer's memory storage device. Alongside the maximum limit for the number of singular values, the maximum truncation error can be used to limit the growth of the network where the truncation error is determined by comparing the Frobenius norm\footnote{Frobenius norm is defined as 
	\begin{eqnarray}
		||M|| = \sqrt{ \sum_{\forall i} |M_{i_1\cdots i_n}|^2}\quad ,\nonumber
	\end{eqnarray}
	where $M$ is a rank-N tensor with indices $ i_{1\cdots n} $.}~\cite{frobenius} of the initial and final tensors. This will allow the network to shrink by discarding the small singular values to satisfy the maximum error condition. One can also achieve a similar effect by explicitly imposing a threshold for the value that singular values can take.
	
	Whilst the DMRG-like updating algorithm reduces the need for hyperparameter optimization, it does not allow an update via SGD algorithms~\cite{stoudenmire2017supervised}. Hence in this study, we propose a combined updating scheme where each epoch starts with the DMRG-like algorithm on a batch and optimizes the network for a limited number of sweeps. This will allow the network to expand with respect to the complexity of the classification problem. Then for other batches, the network is updated using the SGD with the bubbling algorithm, shown in \autoref{fig:keras-like}. 
	
	%%%%%%%%%%%%%%%%%%%%%%%%%%%%%%%%%%%%%%%%%%%%%%%%%%%%%%%%%%%%%%%%
	\section{Top Tagging Through Matrix Product States}\label{sec:toptagging}
	%%%%%%%%%%%%%%%%%%%%%%%%%%%%%%%%%%%%%%%%%%%%%%%%%%%%%%%%%%%%%%%%
	The energy deposits of particles in the electromagnetic and hadronic calorimeters in ATLAS and CMS experiments have long been used to reconstruct the underlying event to understand the physical system better. It has been repeatedly shown that mapping these energy deposits on a modified $\eta-\phi$-plane and analyzing them with a CNN can efficiently discriminate top quark signal over QCD background. In such mapping, each pixel on $\eta-\phi$-plane corresponds to one or more particles depending on their distribution through the detector geometry. A CNN algorithm can pick up the transitionally invariant features of the underlying physics. Due to different particle clusters occurring on the $\eta-\phi$-plane, each pixel is only closely correlated with the close-by pixels. The $\eta-\phi$-plane can be mapped onto an entangled 1D lattice. Since each pixel is only closely correlated with the neighbouring pixels, an MPS can efficiently classify such images. For this study, we will use the preprocessing prescription used in ref.~\cite{Araz:2021wqm} and demonstrate the usage of MPS for top versus QCD jet discrimination. The implementation of this study can be found at \href{https://gitlab.com/jackaraz/tn_classifier}{this link}\footnote{\href{https://gitlab.com/jackaraz/tn_classifier}{https://gitlab.com/jackaraz/tn\_classifier}}.

	%%%%%%%%%%%%%%%%%%%%%%%%%%%%%%%%%%%%%%%%%%%%%%%%%%%%%%%%%%%%%%%%
	\subsection{Dataset \& Preprocessing}\label{sec:preprocess}
	%%%%%%%%%%%%%%%%%%%%%%%%%%%%%%%%%%%%%%%%%%%%%%%%%%%%%%%%%%%%%%%%
	In order to demonstrate the usage of the MPS-classifier, we will use the dataset provided in~\cite{Kasieczka:2019dbj, kasieczka_gregor_2019_2603256}. This dataset consists of top signal and QCD background generated at 14 TeV centre-of-mass energy. The parton level events have been showered in \textsc{Pythia}~8~\cite{Sjostrand:2014zea} and the detector simulation has been obtained via \textsc{Delphes}~3 package~\cite{deFavereau:2013fsa} using the default ATLAS detector card. The so-called fat-jets are reconstructed via \texttt{anti-kT} algorithm~\cite{Cacciari:2008gp} with $R=0.8$ which is employed in \textsc{FastJet}~\cite{Cacciari:2011ma} package. The transverse momentum of these jets are limited to $[550,\ 650]$ GeV range with $|\eta| < 2$. In order to label the dataset, a parton matching with truth level tops have been applied where $\Delta R(j,\ t_{truth})<0.8$ have been required from each event to be considered as signal. The complete dataset includes 1.2 million training, 400,000 validation and test events respectively.
	
	\begin{figure}[!h]
		\centering
		\includegraphics[scale=.435]{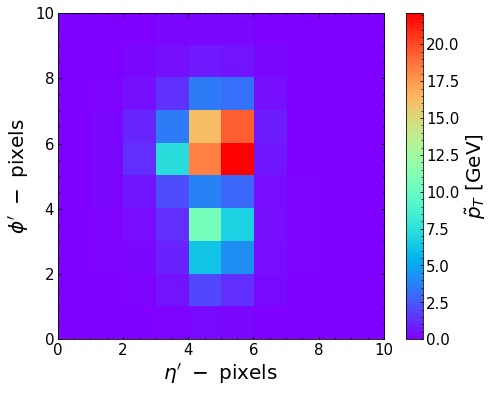}
		\includegraphics[scale=.435]{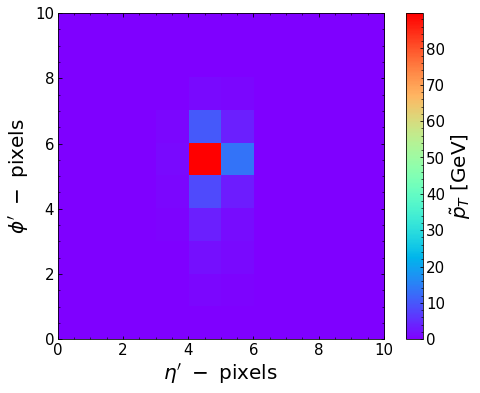}
		\caption{\it The left panel shows the top-signal image on the modified $\eta-\phi$ plane, and the right panel shows the same for the QCD background. The image has been cropped and downsampled from the original by averaging four square pixels; hence the colour of each pixel represents the average transverse momentum of the original four pixels. Each image consists of an average of $5,000$ events. \label{fig:image}}
	\end{figure}

	Following the prescription presented in ref.~\cite{Araz:2021wqm}, the training images are generated by further preprocessing the dataset where each image includes the leading fat-jet (reconstructed as described above). The constituents of the leading fat-jet have been centered with respect to $p_T$ weighted centroid where the jet vector shifted to the centre of $\eta-\phi$-plane. The coordinate system has been shifted to align with the direction of the positive~$\eta$. Furthermore, all the partonic activity has been collected into a $37\times37$ pixelated frame where both $\eta$ and $\phi$ spaning within $[-1.5,\ 1.5]$ range. The lower half of each pixellated image has been flipped to the top if the total $p_T$ of the lower half is greater than the upper half. The same process applied to the left half of the image to ensure that the first quadrant always has the highest $p_T$. In order to simplify the problem, eight pixels from each side of the image have been cropped, and the resulting image has been downsampled by averaging each four-square pixels. \autoref{fig:image} shows the downsampled image where the left panel shows the top signal and the right panel shows the QCD background. Each image has been averaged from 5,000 events.
	
	\begin{figure}[!h]
		\centering
		\includegraphics[scale=1.5]{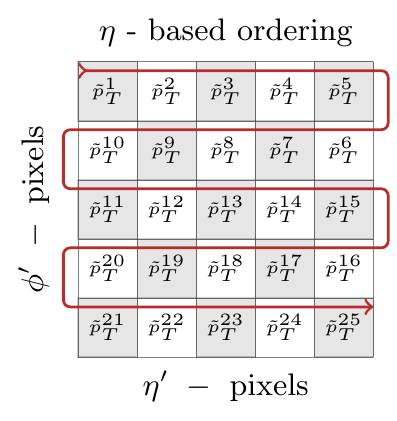}\hspace{1.5cm}
		\includegraphics[scale=1.5]{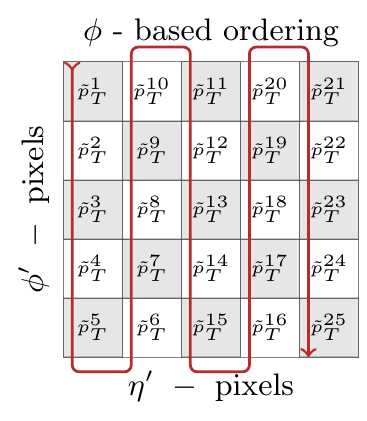}
		\caption{\it The reshaping procedure has been shown to map the image on the 1D lattice. For simplicity, only $5\times 5$ portion of the image has been displayed.  \label{fig:order}}
	\end{figure}
	
	In order to make use of the MPS-classifier, the image data was reshaped to a 1D lattice with an S-shaped mapping, as shown in \autoref{fig:order} where the MPS sites follow the red line where the left panel shows a $\eta$-based ordering, the right panel shows a $ \phi $-based ordering. Note that this representative figure is shrunk to $5\times 5$ just for visual simplicity. This S-shaped reshaping ensures that the alternating edges of the image are in proximity to maximize the entanglement\footnote{Ref.~\cite{selvan2020tensor} discusses other possible preprocessing methods which might improve the classification by reordering the lattice to make sure all the neighbouring pixels are at maximally proximity.}. After normalizing images by 650 GeV, each pixel has been mapped onto a hypersphere via
	\begin{eqnarray}
		\phi^{p_i}(\tilde{p}^i_T) = \sqrt{\binom{D-1}{p_i - 1}}\ \cos^{D-p_i}\left(\tilde{p}^i_T \frac{\pi}{2}\right)\ \sin^{p_i - 1} \left(\tilde{p}^i_T \frac{\pi}{2} \right)\ ,\quad p_i \in 1,\ldots, D\ ,\label{eq:hypersphere}
	\end{eqnarray}
	  where in two dimensions ($ D=2 $) $\phi^{p_i}(\tilde{p}^i_T) $ reduces to $ [\cos(\tilde{p}^i_T \pi /2),\ \sin(\tilde{p}^i_T \pi /2)]$. Here $\tilde{p}^i_T$ corresponds to the modified transverse momentum within the pixel $ i $.
	
	%%%%%%%%%%%%%%%%%%%%%%%%%%%%%%%%%%%%%%%%%%%%%%%%%%%%%%%%%%%%%%%%
	\subsection{Network architecture \& training}\label{sec:training}
	%%%%%%%%%%%%%%%%%%%%%%%%%%%%%%%%%%%%%%%%%%%%%%%%%%%%%%%%%%%%%%%%
	To capture the MPS-classifier's capability, we will demonstrate our results for different training procedures and compare the outcomes to the state-of-the-art CNN results. Our MPS classifier relies on \textsc{TensorFlow} version 2.4.1~\cite{tensorflow2015-whitepaper, DBLP:journals/corr/AbadiBCCDDDGIIK16} and \textsc{TensorNetwork} version 0.4.3~\cite{Roberts:2019qim} and the CNN relies on \textsc{Keras} package~\cite{chollet2015keras} embeded in \textsc{TensorFlow}.
	
	As a baseline, we choose to use a CNN architecture presented in ref.~\cite{Araz:2021wqm}. The architecture takes in $10\times10$ image pixels with a batch size of 128. The images are then passed through a convolutional layer with eight features and four stride pixels, including zero padding. After a batch normalization layer, the latent image is passed through a max-pooling layer where the size of the image is dropped to $5\times 5$ pixels. Then the resulting latent image has been flattened and passed through a fully connected layer with sixteen nodes. For each layer \texttt{ReLU} activation function has been used. The network has been trained for 500 epochs and learning rate has been decayed from $0.01$ every $20$ epochs if validation loss has not been improved. In all the networks presented below, we used the cross-entropy loss function,
	\begin{eqnarray}
		\mathcal{L} = - \frac{1}{N}\sum_{x\in\mathbf{x}^N} y^{truth}\ \log \left(\hat{y}\right) \ , \nonumber
	\end{eqnarray}
	where $N$ represents the number of events in a batch. 
	
	\begin{figure}[!h]
		\centering
		\includegraphics[scale=.4]{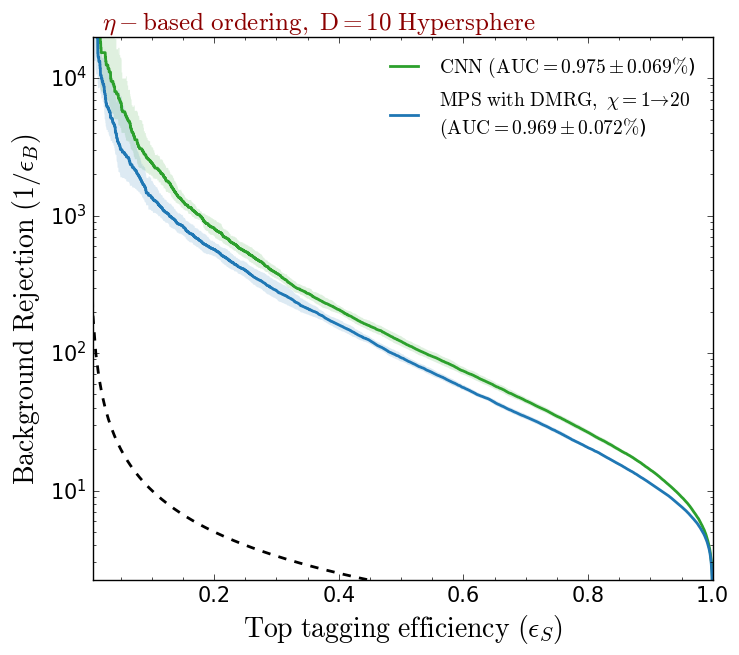}
		\includegraphics[scale=.4]{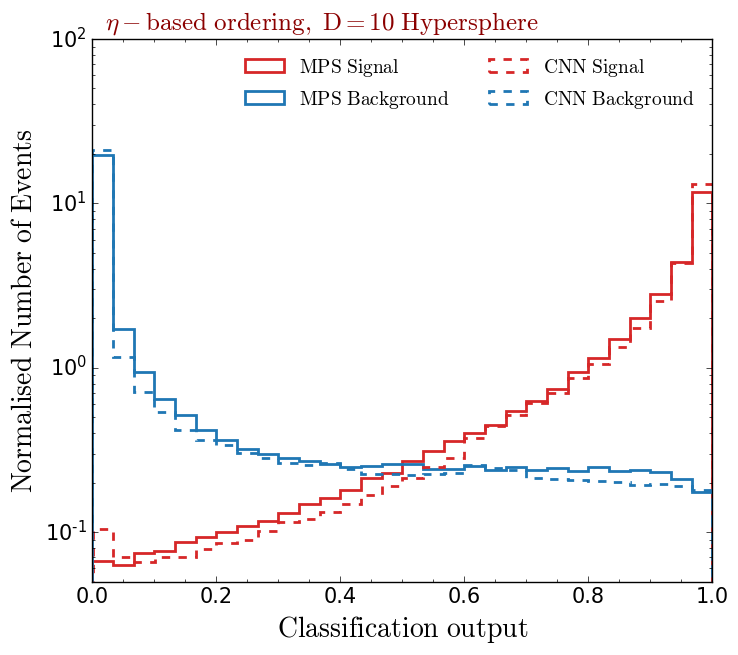}
		\includegraphics[scale=.425]{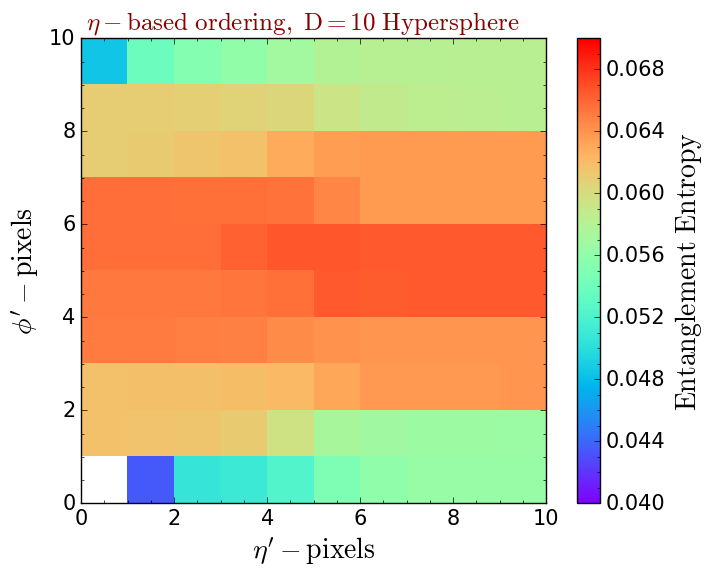}
		\includegraphics[scale=.38]{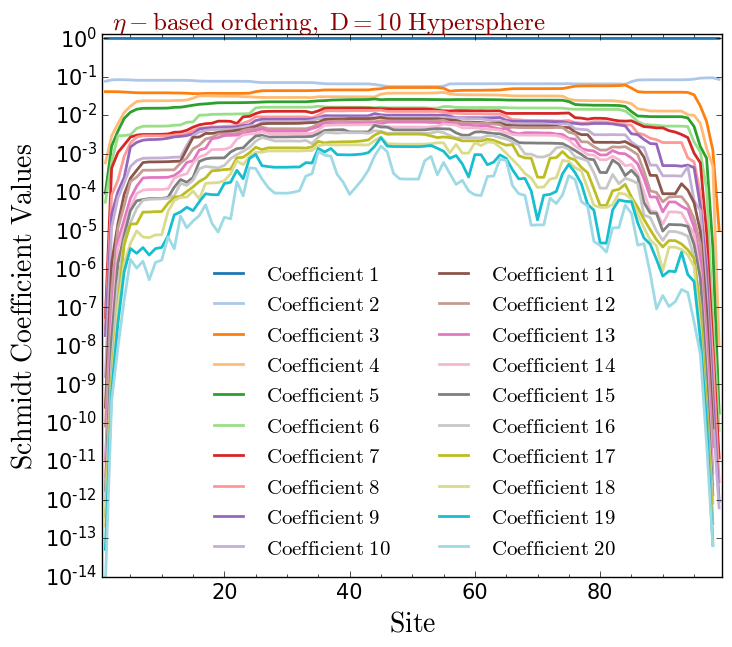}
		\caption{\it The Upper right panel shows the receiver operating characteristic curve for CNN and MPS classifiers correspond to green and blue curves, respectively, and the left panel shows the corresponding classifier output. The statistical uncertainty has been measured by dividing the test set into batches of 50,000 events. The dashed black curve corresponds to random choice. The bottom left panel shows the entanglement entropy mapped on $\phi-\eta$ frame, and the right bottom panel shows the respective Schmidt coefficient distribution for each MPS site.}\label{fig:d10_results}
	\end{figure}
	
	In order to compare with the CNN, we first mapped our $\eta$-based ordered images onto a 10D hypersphere as shown in \autoref{eq:hypersphere}. Then the network has been trained via the DMRG algorithm. We initially assumed no entanglement between the lattice nodes within the quantum state, and allowed it to expand up to 20 auxiliary dimensions. In order to avoid getting stuck in a local minimum, the growth of the network has been done gradually where in the first epoch the network has been allowed to expand up to 10 auxiliary dimensions and then each epoch this value increased by 1.5 times its size, eventually reaching 20. Without such gradual growth, we observed that the network could not reach the presented stage within a limited amount of epochs. The MPS-classifier has been trained for 200 epochs with a batch size of 10,000 events. The learning rate was chosen to be 0.0001 and decayed to its half every 20 epochs if no improvement has been observed in the validation loss. Upper left panel of \autoref{fig:d10_results} shows the receiver operating characteristic (ROC) curve for MPS-classifier (blue) versus CNN (green). The statistical uncertainty has been calculated by splitting the test set into batches of 50,000 events, where shaded area around each curve represents one standard deviation. Similarly the right panel of \autoref{fig:d10_results} shows the classification output of both networks where the blue solid (dashed) line represents the background for MPS-classifier (CNN) and the red solid (dashed) line represents the signal. Despite the large batch size, the MPS-classifier has achieved very similar prediction accuracy compared to the CNN classifier, where their corresponding AUC values differ by only 0.62\%.
	
	 The bottom left panel of \autoref{fig:d10_results} shows the entanglement entropy of the MPS classifier and the right panel shows the Schmidt coefficient values, calculated as shown in  eqs.~\eqref{eq:schmidt_decomp} and \eqref{eq:entanglement_entropy}. One can immediately observe that the entanglement entropy follows the pixel unfolding procedure shown in the left panel of the \autoref{fig:order}. Additionally, entropy follows a grouped pattern where instead of a sharp increase or decrease, it preserves similar entropy among around 10 pixels before changing the entropy value. This shows that not all pixels are equally valuable, where one can avoid using the pixels with the same entropy values hence shows a path for feature compression. One can observe that the network takes the same entropy value mainly on the left or right side of the image and changes at the centre. This is because most of the information is at the centre of the image and sides are generally empty; hence information has been propagated through the pixels that do not posses additional information. Note that the bottom-left pixel is empty simply because each node is entangled with the node to its right (or left); due to non-periodic boundaries in MPS the last node does not have any connection on its right. In the right panel, we present the Schmidt value distribution in each node, used to calculate the entanglement entropy. In this particular example, we observe that Schmidt values can be smaller than $10^{-14}$. As can be seen from the plot, the number of Schmidt coefficients decreases towards the edges of the MPS. This is due to the canonical structure of the MPS with non-periodic boundaries, where the number of auxiliary dimensions increases towards the centre. Each auxiliary dimension would be the same for an MPS with periodic boundaries, assuming no Schmidt value trimming has been imposed.
	
\subsubsection{Feature space and network compression}
\label{sec:notrain}
	
	Whilst a large number of trainable parameters gives the network chance to sample various possibilities to achieve the optimisation task at hand, it also leads to a vast loss hypersurface to explore. It can be challenging to explore and find a global minimum in such an ample parameter space, even with advanced optimisation algorithms. Hence, it is vital to optimise a given hypothesis's hyperparameters to achieve faster, if not better, convergence to a global minimum. Beyond the convergence, this will also lead to a quicker inference after the training.
	
	\begin{figure}[!h]
		\centering
		\includegraphics[scale=.38]{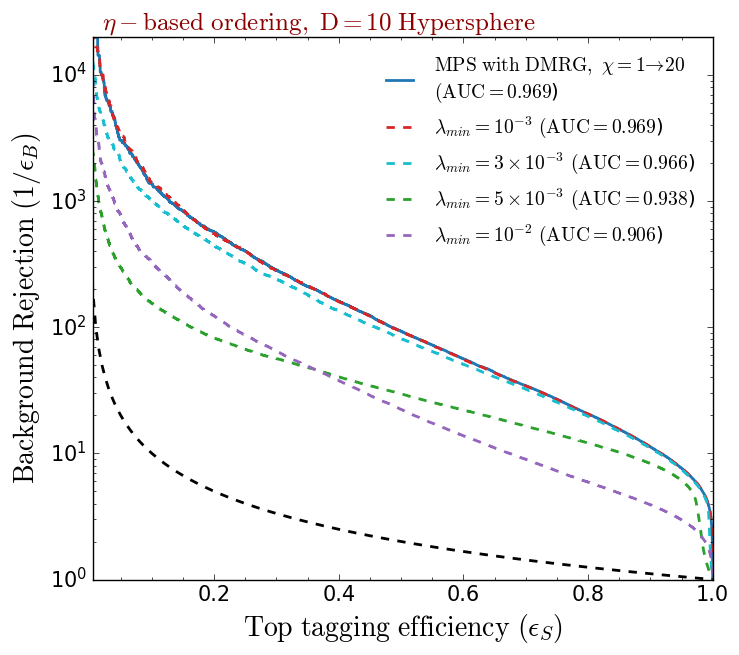}
		\includegraphics[scale=.38]{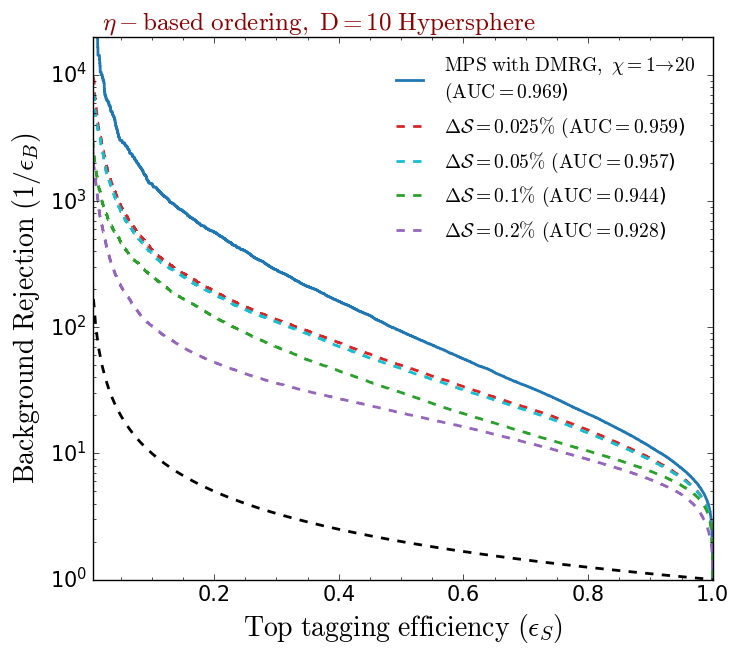}
		\includegraphics[scale=.38]{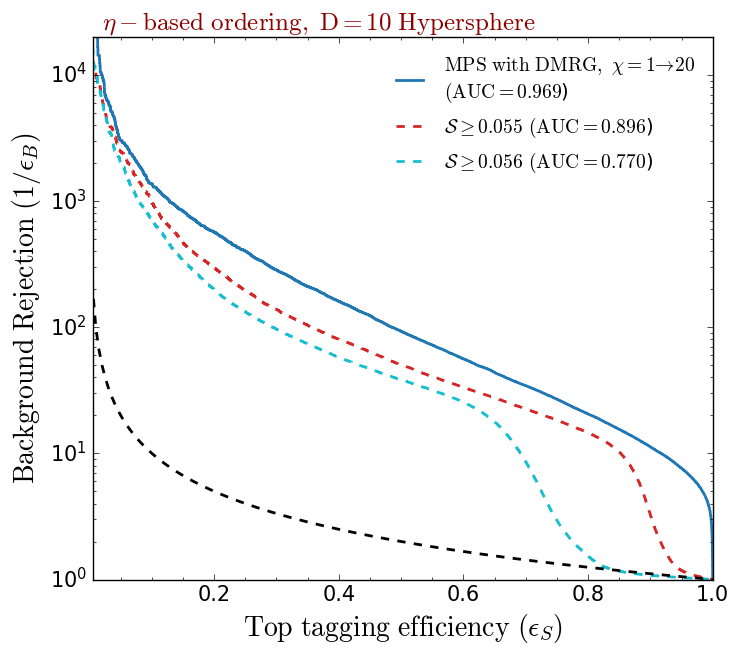}
		\caption{\it Three types of network compression after training have been presented. The top left panel shows the ROC curve after introducing certain thresholds on the Schmidt values, the top right panel shows the same for entropy change based network compression (see the text for details), and the bottom panel shows the ROC curve for the compressed network with entropy thresholds.}\label{fig:compression}
	\end{figure}
	
	The entanglement entropy and the Schmidt coefficients are especially valuable to provide a path towards network compression. One can use them to design a compressed feature space or limit the network size to achieve faster predictions. \autoref{fig:compression} shows three different methodologies that we adapted to reduce the size of the network while quantifying the effects of the compression on the precision of its network output. On the upper left panel, we show the effect of applying the minimum Schmidt value threshold ($\lambda_{min}$) on the network after the training. Following the bottom right panel of \autoref{fig:d10_results}, we applied $\lambda_{min}=10^{-3},\ 3\times10^{-3},\ 5\times10^{-3}$ and $10^{-2}$ shown with dashed red, cyan, green and purple ROC curves respectively. Each step number of trainable parameters dropped from 390500 to 204310, 91690, 32990 and 18020, respectively. Although there is a considerable 76\% drop in the number of trainable parameters, we do not observe a significant change in the precision until we reach $\lambda_{min}=3\times10^{-3}$.

	On the top right panel of \autoref{fig:compression}, we applied an entanglement entropy-based compression of the network. The change in entanglement entropy signals the contribution of information that can be used for the classification task. If $\Delta \mathcal{S}$ does not change along the different sites, one can eliminate the features with the same entanglement entropy while maintaining a similar precision. Hence we binned the entanglement entropy values with respect to the change in entropy denoted by $\Delta \mathcal{S}$ where the entropy change between the first and last element in the bin corresponds to given $\Delta \mathcal{S}$ value. Among these bins, we chose the nodes with the most significant entanglement and formed a new network. We present the ROC curves with $\Delta \mathcal{S}=0.025\%,\ 0.05\%,\ 0.1\%$ and $0.2\%$ following red, cyan, green and purple ROC curves which reduced the network size to 83, 75, 64 and 54 nodes. 
	
	Finally, at the bottom panel of \autoref{fig:compression} we adopted an entropy-threshold-based compression where only the nodes with entropy greater than indicated have remained in the new network. We present the results from the network with $\mathcal{S}\geq 0.055$ and $0.056$ presented with red and cyan ROC curves. These thresholds dropped the number of sites in the network to 95 and 94, respectively.
	
	Comparing all three compression methods presented reveals the nature of the DMRG training algorithm. This method spans the geometrical structure of the feature space and attempts to understand the entanglement in the 1D lattice wave function. The number of auxiliary dimensions shows the length of entanglement in each node when it is reduced, as shown in the upper left panel of \autoref{fig:compression},  the sites becomes less entangled, and the outcome becomes less precise. Entanglement based restriction shows the location of the stored information where the information regarding the 3-prong structure of top has been spread towards the outskirts of the MPS; hence removing these nodes degrades the precision of the network at high tagging efficiency region. The information regarding the dipole structure, on the other hand, has been preserved at the centre of the network, where the entanglement entropy is at its larges. While removing the nodes with low entanglement on the edges does not affect the dipole substructure information, the $\Delta S$ method effectively removes the nodes in each region, including the region with the highest entropy, which effectively reduces the precision in low top tagging efficiency. This exercise reveals that the DMRG algorithm is capable of learning the geometrical structure of the given data.
	
	\begin{figure}[!h]
		\centering
		\includegraphics[scale=.45]{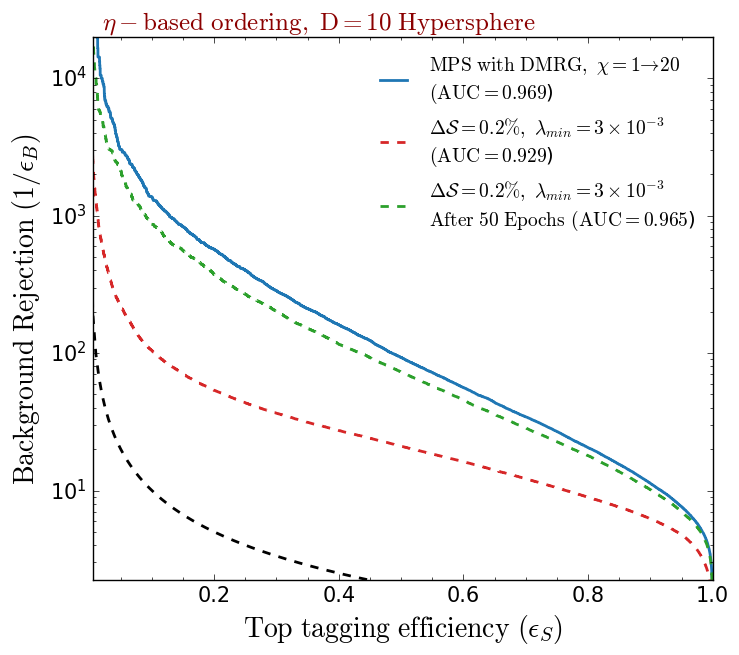}
		\caption{\it ROC curves to compare the original network (solid blue), compressed network (dashed red) and the compressed network after 50 epochs (dashed green). The dashed black curve shows the random choice.} \label{fig:compressed_train}
	\end{figure}

	Combining the compression methods of $\Delta \mathcal{S}$ and $\lambda_\mathrm{min}$, we formed a new network and retrained it from scratch. \autoref{fig:compressed_train} shows the ROC comparison between the original network, compressed version and the trained version. Initially, the original network (shown with the solid blue curve) compressed by imposing $\Delta \mathcal{S} = 0.2\%$ and limiting the Schmidt coefficient values above $3\times10^{-3}$ (shown with the dashed red curve). This compression limited the network to only 54 sites (or pixels) and 43410 parameters. Then we trained this network for another 50 epochs with a learning rate starting at $10^{-4}$ and decaying as before. During the training, we set a rigit Schmidt threshold at $3\times10^{-3}$. This resulted in further compression of the network to 34160 trainable parameters—the ROC curve after 50 epochs presented with the dashed green curve.  Although the number of parameters of the new network is only 8\% of the original one, it managed to achieve very similar precision. This shows that the entanglement entropy holds the crucial information behind what the network learns.
	
	\subsubsection{Training algorithm comparison}
	
	The MPS-classifier's prediction can also be calculated using the SGD algorithm. Note that, although we also use a mini-batch SGD type backpropagation in the DMRG algorithm, henceforth SGD acronym will only refer to a neural network type of backpropagation where all the parameters are updated simultaneously using a sophisticated algorithm such as \texttt{Adam}~\cite{Kingma2014AdamAM}. In order to compare these algorithms in a more straightforward framework, we reduced the hypersphere mapping to 2D and investigated the effects of both $\eta$ and $\phi$ based ordering for sake of completeness. Although $\eta$-based ordering has been shown to have a significant impact on the classification results, the pixels do not possess a correlation in the vertical axis, which might introduce different properties. Whilst DMRG training has been done as before, SGD training has been achieved by the \texttt{Adam} algorithm~\cite{Kingma2014AdamAM}. After the contraction of the MPS-classifier, the gradients of each node in the MPS has been calculated with respect to the objective function. We observed that the normalised gradient tensors lead to more stable results. Since the SGD algorithm cannot adapt the network with respect to the complexity of the problem, we initialised the network with ten bond dimensions per node, and the network has been canonicalised to limit the number of trainable parameters prior to the training, in agreement with the DMRG construction. The rest of the training hyperparameters are set equal to the DMRG algorithm. Furthermore, to harvest the ability of both algorithms, we set up a DMRG+SGD algorithm where each epoch applied the DMRG algorithm to the first batch for tree sweeps, and the rest of the batches are trained with SGD. This combination gave the network the ability to adjust its size with respect to the complexity of the problem. Hence, we assumed no entanglement in the network prior to the training, i.e. the network can grow without restrictions except from an upper limit on the bond dimensions. 	
	\begin{figure}[!h]
	\centering
	\includegraphics[scale=.4]{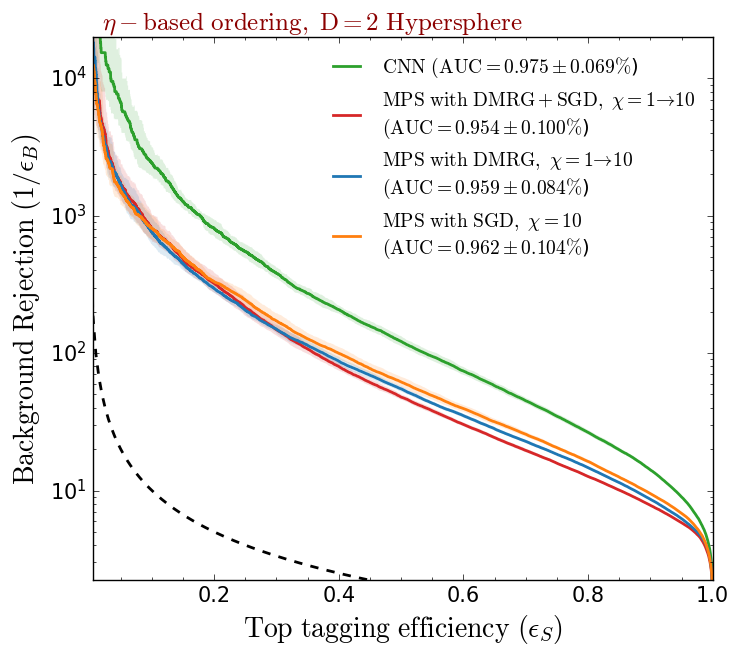}\hspace{-1mm}
	\includegraphics[scale=.4]{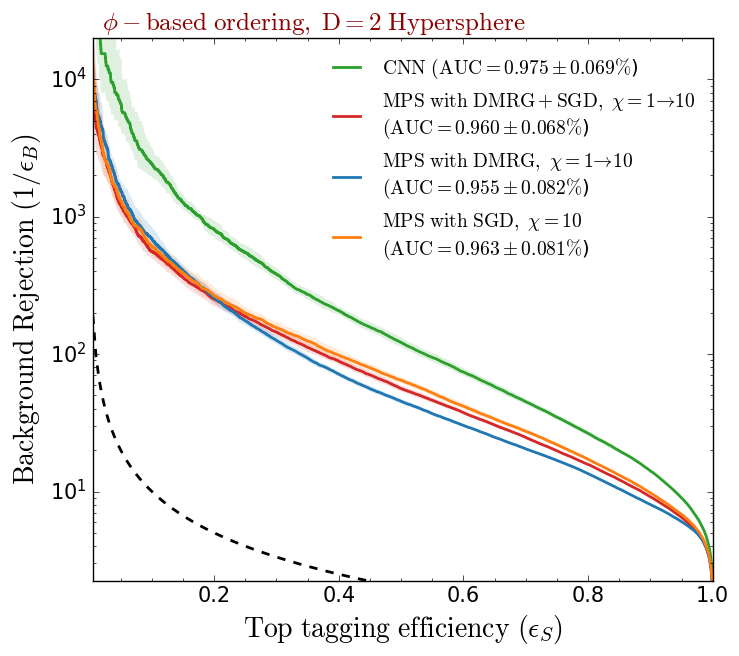}
	\caption{\it The left panel shows ROC curves for convolutional and MPS network classifiers with $\eta$-based ordering, and the right panel shows the same for $\phi$-based ordering. The green, red, blue and orange curves represent CNN, MPS-classifier trained with DMRG+SGD algorithm,  DMRG algorithm, and SGD algorithm. The shaded area shows the statistical uncertainty defined same as \autoref{fig:d10_results}. The dashed black curve represents the random choice.  \label{fig:roc}}
	\end{figure}	
	
	\autoref{fig:roc} shows the ROC curve for all the networks and training algorithms presented above, where blue, red and orange curves represent DMRG, DMRG+SGD and SGD algorithms. The left panel shows results for  $\eta$-based ordered MPS, and the right panel shows the same for $\phi$-based ordered MPS. Compared to the 10D hypersphere, we observe a slight decrease in the generality of MPS classification results; however, we only observe slight differences between the results of the different training algorithms. All algorithms seem to be able to discriminate the dipole type substructure with equal performance, and the difference seems to be more towards discriminating 3-prong substructure. This directly shows the effect of the local entanglement in MPS which is directly linked to the lattice order. 
	
	\begin{figure}[!h]
		\centering
		\includegraphics[scale=.4]{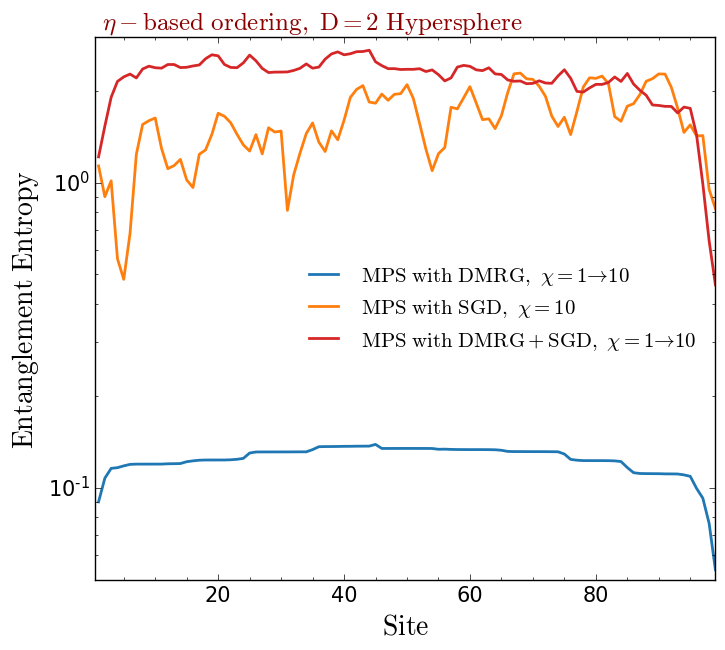}
		\includegraphics[scale=.4]{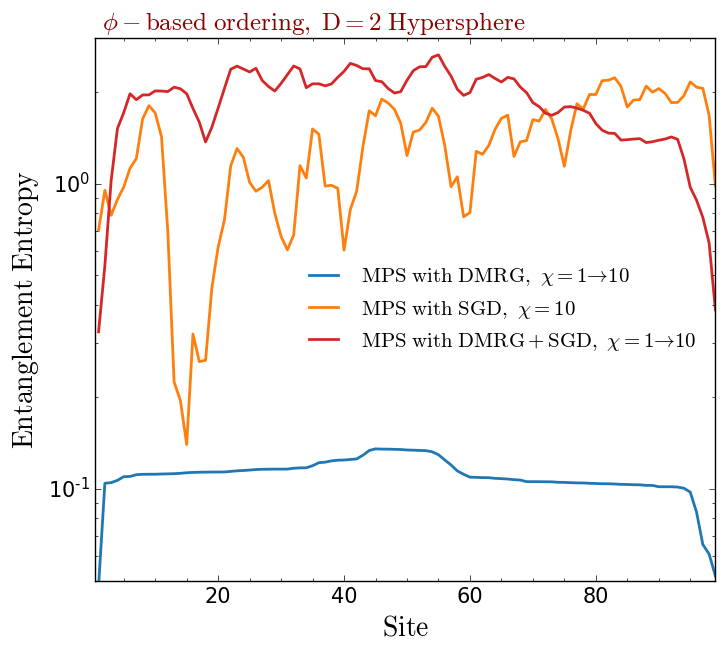}
		\caption{\it The left panel shows the entanglement entropy for $\eta$-based ordered MPS where blue, red, and orange curves correspond to the MPS trained by DMRG, DMRG+SGD and SGD algorithms, respectively. The right panel shows the same for $\phi$-based ordered MPS. \label{fig:entropy}}
	\end{figure}

\begin{figure}[!h]
		\centering
		\includegraphics[scale=.4]{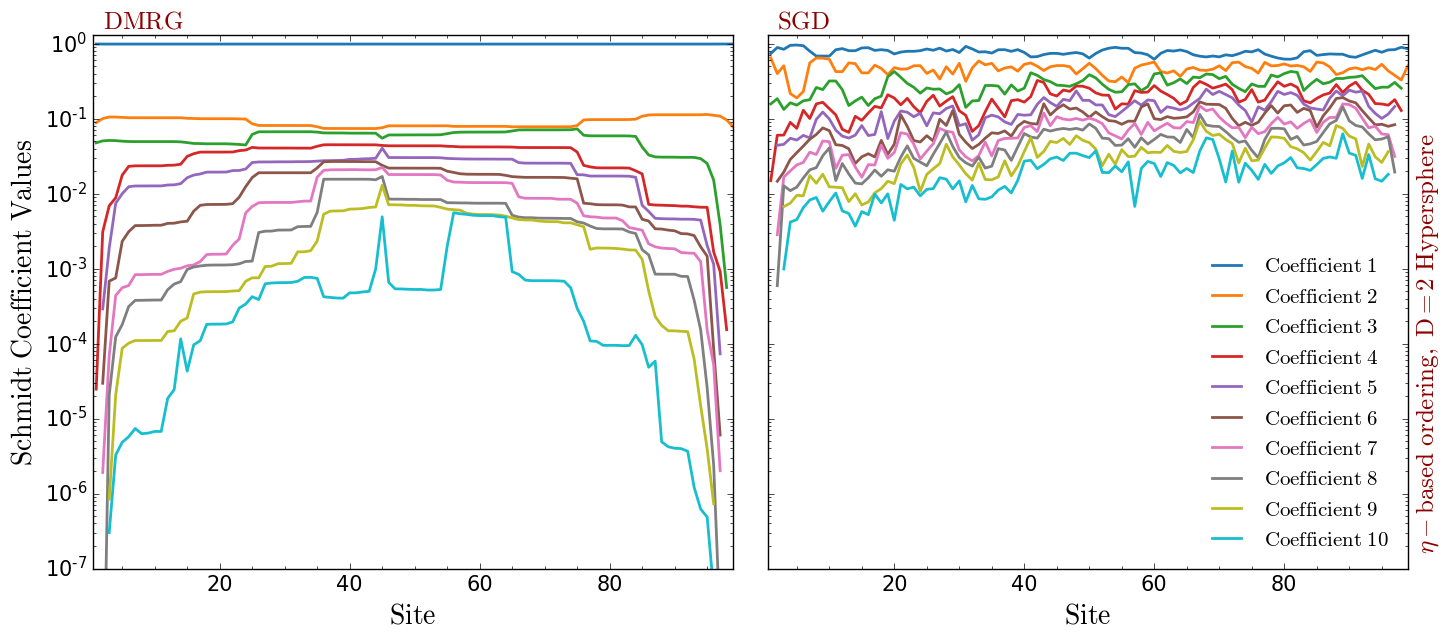}
		\includegraphics[scale=.4]{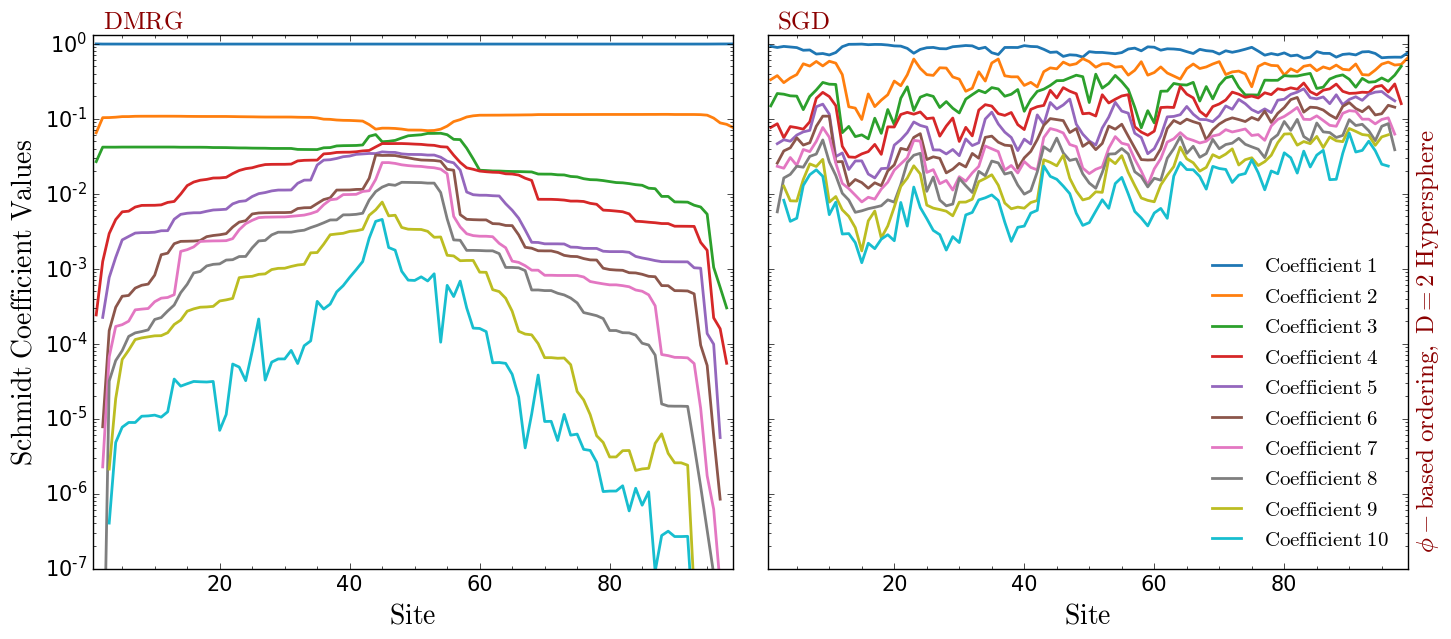}
		\caption{\it The upper two panels show the Schmidt Coefficient distributions for $\eta$-based ordered MPS where the left panel used for MPS trained by DMRG and the right panel for SGD algorithms. The bottom two panels show the same for $\phi$-based ordered MPS.  \label{fig:schmidt}}
	\end{figure}

	Although there is no significant difference in the classification results, we observe large variations in what the training algorithm captures. \autoref{fig:entropy} shows the entanglement entropy per site for each algorithm following the same colour scheme as before, and the left (right) panel shows the entropy for $\eta(\phi)$-based ordered image. As observed in \autoref{fig:d10_results}, the DMRG algorithm is trying to set seemingly similar entropy distribution for each site where its peaking at the centre of the image, the area corresponding to the most hadronic activity. Compared to $\eta$-based ordering, $\phi$-based ordering seems to capture two prongs in the vertical axis, hence the increased entropy between sites 40 and 60. Compared to the DMRG, on the other hand, the SGD algorithm seems to capture much more information which also reflects to the DMRG+SGD algorithm. 
	This approach can capture larger entropy since SGD attempts to increase the degrees of freedom of the network, it does not capture irrelevant pixels. Although this is a desirable feature in terms of classification performance, it limits the interpretability of the network. The SGD algorithm is susceptible to the rare energy fluctuations on the edges of the image; however, the DMRG algorithm seems to learn not to rely on non-frequent data. Whilst this does not affect the performance between DMRG and SGD algorithms significantly in this particular example, the SGD algorithm might be superior to DMRG for the cases with large fluctuations in the data.
	This effect is also captured in Schmidt coefficients observed in two algorithms; upper and bottom part of the \autoref{fig:schmidt} shows the Schmidt coefficient distribution for $\eta$ and $\phi$-based ordering. The left panel of both images shows it for the DMRG algorithm, where the right panel shows for the SGD algorithm. For both orderings, the SGD algorithm seems to have similar Schmidt coefficient distribution where all the singular values agregate above $\sim10^{-2}$. The DMRG algorithm, on the other hand, seems to preserve the information embedded by the geometrical structure of the ordering where the singular values are peaking at the centre of the image, and the nodes are less and less expressive at the beginning and the end of the network. This vital information holds the key towards compressing the network and eliminating the less expressive connections between nodes as shown before. The DMRG algorithm not only learns data to achieve the best possible minimum for the loss function but also captures the physical properties of the data reflected in the entanglement entropy and Schmidt distribution. This also shows that in a less correlated feature space SGD would be expected to perform better than DMRG.
	
	Comparing ordering schemes shows a couple of main differences between them. Whilst the entanglement entropy captured by the DMRG algorithm is more or less the same for both orderings, the entropy change ($\Delta \mathcal{S}$) in $\phi$--ordering is mostly effective between sites $ 40-60 $. This allows for more significant feature space compression with respect to $\eta$--based ordering. Similarly, Schmidt values for DMRG, presented in \autoref{fig:schmidt}, shows that $\phi$--based ordering requires less precision to achieve the same accuracy as $\eta$--based ordering. This is due to the alignment of the image where most of the hadronic activity is happening between $ 0-6 $ $\eta^\prime$ pixels, see \autoref{fig:image}, which is captured by the DMRG algorithm by decreasing the Schmidt values after the $50^{\rm th}$ site. Hence $\phi$--based ordering can be used to achieve more extensive compression of both feature space and network without loss of generality. 
		%This effect is also visible in the SGD approach; while Schmidt values in $\eta$--based ordering reaching up to $10^{-2}$, in $\phi$--based ordering, they are reaching up to $10^{-3}$. }

	\begin{figure}[!h]
		\centering
		\includegraphics[scale=.43]{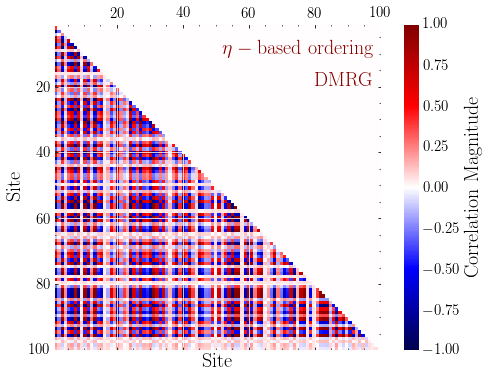}
		\includegraphics[scale=.43]{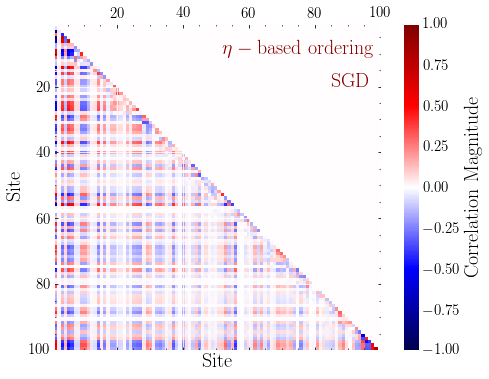}
		\caption{\it The left panel shows the correlation matrix, with respect to $ \sigma_z $, of the MPS sites which has trained via the DMRG algorithm, and the right panel shows the same for the MPS trained with SGD. \label{fig:corr}}
	\end{figure}

	As mentioned before, two-site correlations of the sites can be helpful to unfold the properties of these algorithms even further. \autoref{fig:corr} shows the normalised correlations of the MPS sites that trained with the DMRG algorithm (on the left) and SGD algorithm (on the right) with respect to the third Pauli matrix, $\sigma_z$. Due to the nature of the DMRG algorithm, we observe many (anti-)correlated pixels, which essentially coincides with the entanglement entropy distribution. As shown in ref.~\cite{trenti2020quantuminspired}, this information can be used alongside entanglement entropy to further compress the network. However, the network that is trained via SGD hasn't shown any significant correlation between pixels compared to DMRG-based training, where the largest (anti-)correlation reaching up to $\pm 0.8$. This shows further support for the claim of the nature of these algorithms; as mentioned before, the DMRG algorithm is actively decreasing the degrees of freedom by iteratively selecting the most prominent features by correlating them while the SGD algorithm is trying to actively increase the expressibility of the network. Although this renders the results of the SGD algorithm hard to interpret, it might be beneficial in challenging optimisation problems.

	During the training, we did not observe overtraining due to the large batch size. The overtraining has only been observed when the batch size reduced below 500. The SGD algorithm, seems to reach a well-trained level at around 200 epochs; however, the DMRG algorithm requires more extended training time to reach that level. Hence, the results presented for the DMRG algorithm are all somewhat undertrained. On the other hand, the DMRG algorithm reaches a good discrimination level within couple epochs, and during the rest of the training it only slightly decreases the loss value which has also been captured in other studies~\cite{stoudenmire2017supervised, efthymiou2019tensornetwork}. Hence, the DMRG algorithm is suitable to achieve decent results with little training time, but it requires a longer training time to reach its full potential.
	
	Despite the interpretability, TN based applications can come with a significant toll on computation time. To assess the timing, we measured the elapsed time for bond dimensions 10 and 20 with 2D data embedding. As above, batch size of $10,000$ events has been used for each training algorithm. The benchmarks have been collected for NVIDIA Tesla V100 16GB GPU, where the algorithms are parallelized only with respect to the training samples. A complete sweep (starting from the left node and ending at the left node) for the DMRG algorithm with 10 (20) bond dimensions and 100 sites have been measured to take 6 (6.02) seconds on average after complete compilation of the algorithm. The calculation of a single batch on the SGD algorithm, on the other hand, took 0.46 (0.47) seconds for a maximum bond dimension of 10 (20). Note that although these benchmarks are from fully parallelized algorithms, it is possible to make them more efficient. For instance, the DMRG algorithm presented in this study has been optimized to preserve the canonical form of the MPS throughout the training, which means that each update has only been applied to the centre-of-orthogonality tensor. However, it is possible to write a much faster algorithm by contracting adjacent tensor pairs and updating them simultaneously. This will allow more NN-like backpropagation but will not preserve the canonical form of the network. Similarly, the SGD algorithm presented in this study does not use the efficient {\sc Keras} integration in order to manipulate network structure and use normalized gradients to update the network. More efficient {\sc Keras}-based TN classifiers can be found in {\sc TensorNetwork}~\cite{Roberts:2019qim} library\footnote{For similar implementations also see refs.~\cite{torchmps, JMLR:v21:18-008, itensor, Suess2017, 10.21468/SciPostPhysLectNotes.5}.}.

	%%%%%%%%%%%%%%%%%%%%%%%%%%%%%%%%%%%%%%%%%%%%%%%%%%%%%%%%%%%%%%%%
	\section{Conclusion}\label{sec:conclusion}
	%%%%%%%%%%%%%%%%%%%%%%%%%%%%%%%%%%%%%%%%%%%%%%%%%%%%%%%%%%%%%%%%
	
	Tensor Networks are non-trivial representations of high-dimensional tensors. They have developed into powerful numerical tools to perform highly sophisticated calculations of complex quantum systems. In the context of machine learning Tensor Networks have been shown to be ideal vehicles to connect quantum mechanical concepts to machine learning techniques, thereby facilitating an improved interpretability of neural networks.
	
	In this study, we applied specific TNs, i.e. Matrix Product States, to classify LHC pseudo-data as top quark or QCD-induced processes. We compared state-of-the-art convolutional network-based tagging algorithms to such MPS and showed that it is possible to achieve similar results to the CNN, while gaining a deeper insight into how and what the network learns. 

	The DMRG algorithm has been designed to automatically adapt the network architecture depending on the complexity of the problem at hand, which reduces the need for hyperparameter optimization. We showed that whilst SGD can extract more information from the network in terms of entanglement entropy and two-site correlations of the pixels, DMRG learns the geometrical structure of the data and the correlations between the sites on the 1D lattice. Whereas no significant performance difference has been observed between the two optimization algorithm, DMRG led to more interpretable results. Despite the fact that the MPS only captures one dimensional correlations between pixels it can achieve similar precision as the state-of-the-art CNN classifier.
	
	We also proposed an adaptable algorithm for training MPS by merging DMRG and SGD algorithms within one training sequence. This allowed us to harvest the adaptability of DMRG and the efficiency of the SGD algorithm. However, we observed that since the goal of these algorithms is entirely different, one needs to be mindful while training the network. While the SGD algorithm aims to increase the individual expressivity of each node by increasing the degrees of freedom to achieve high-performance classification, DMRG is trying to reduce the degrees of freedom of the neighbouring nodes to construct a locally entangled network. Hence, the fraction of DMRG sweeps during the training becomes a crucial hyper-parameter where if not adjusted correctly, the network performance can be less than the network separately trained by DMRG or SGD.

	Further, using entanglement entropy, one can devise algorithms to effectively compress the network to reduce the decision making time during prediction, which can be beneficial for experimental analyses~\cite{trenti2020quantuminspired}. Additionally, since the MPS only expresses local entanglement, different pixel reordering schemes are very effective to increase the classification power~\cite{selvan2020tensor}. It has also been shown that MPS can be used to pre-train the Quantum Machine Learning (QML) networks to boost the convergence of the algorithm~\cite{dborin2021matrix}. One shortcoming of the MPS approach is that it cannot fully represent the data contained in the event image, as the pixel have to be represented as a 1D latticised chain. However, such limitation can be mitigated by using 2D tensor networks such as Projected Entangled Pair States (PEPS) to exploit more information from an image~\cite{Cheng_2021} or tree tensor network structure can be used to embed non-trivial connections between the features~\cite{trenti2020quantuminspired}. Such classical methods based on TNs can also be used to estimate the limitations of the algorithms in QML applications~\cite{Zhou_2020, 10.3389/fphy.2020.586374}.

	The development of novel, yet explainable, data analysis methods is of crucial importance for upcoming LHC runs, in particular during its high-luminosity phase. In the context of high-energy physics Tensor Networks are rather unexplored techniques that show promising properties which can complement or even replace existing machine learning approaches. 

	%%%%%%%%%%%%%%%%%%%%%%%%%%%%%%%%%%%%%%%%%%%%%%%%%%%%%%%%%%%%%%%%
	\bibliography{bibliography}
	%%%%%%%%%%%%%%%%%%%%%%%%%%%%%%%%%%%%%%%%%%%%%%%%%%%%%%%%%%%%%%%%
\end{document}